\documentclass[linenumbers,preprint,showkeys,preprintnumbers,amsmath,amssymb,superscriptaddress]{revtex4}
\usepackage{amsmath,amssymb,bm,bbm,eucal,graphicx}

\usepackage{graphicx}
\usepackage{graphics}
\usepackage{subfigure}
\usepackage{natbib}
\usepackage{dcolumn}
\usepackage{bm}
\usepackage{txfonts}
\usepackage{color}

\begin{document}

\centerline{\Large Coevolution of Cooperation and Partner Rewiring}
\centerline{\Large Range in Spatial Social Networks}

{\small {\vskip 12pt \centerline{Tommy Khoo$^{1}$, Feng Fu$^{1,2,*}$, \& Scott Pauls$^{1,**}$}

\begin{center}
$^1$ Department of Mathematics, Dartmouth College, Hanover, NH 03755, USA \\
$^2$ Department of Biomedical Data Science, Geisel School of Medicine, \\ Dartmouth College, Hanover, NH 03755, USA
\end{center}
}}

\begin{center}
[$^*$] Feng.Fu@dartmouth.edu, [$^{**}$] Scott.D.Pauls@dartmouth.edu
\end{center}

\vskip 30pt

\begin{minipage}{142mm}
\begin{flushleft}

{\textbf{Manuscript information:}}\, 39 pages (including figure captions); 4 figures; 14 supplementary figures; \\
\end{flushleft}
\end{minipage}

\clearpage
{
{\bf Abstract:} \, 
In recent years, there has been growing interest in the study of coevolutionary games on networks. Despite much progress, little attention has been paid to \emph{spatially} embedded networks, where the underlying geographic distance, rather than the graph distance, is an important and relevant aspect of the partner rewiring process. It thus remains largely unclear how individual partner rewiring range preference, local vs. global, emerges and affects cooperation. Here we explicitly address this issue using a coevolutionary model of cooperation and partner rewiring range preference in spatially embedded social networks. In contrast to local rewiring, global rewiring has no distance restriction but incurs a one-time cost upon establishing any long range link. We find that under a wide range of model parameters, global partner switching preference can coevolve with cooperation. Moreover, the resulting partner network is highly degree-heterogeneous with small average shortest path length while maintaining high clustering, thereby possessing small-world properties. We also discover an optimum availability of reputation information for the emergence of global cooperators, who form distant partnerships at a cost to themselves. From the coevolutionary perspective, our work may help explain the ubiquity of small-world topologies arising alongside cooperation in the real world.
}

\clearpage

\section{Introduction}

We are surrounded by a remarkable living world which is constantly being shaped by evolutionary dynamics \cite{nowakBook}. Ranging from multicellular organisms \cite{MichodHeridity01}, to animal groups \cite{Clutton-BrockScience01}, to human societies \cite{ApicellaNature12}, cooperation among individuals is indispensable to achieve such high levels of organizational complexity. However, cooperation cannot be taken for granted, since cooperators incur a cost to benefit others, and are thus prone to exploitation by selfish individuals \cite{HardinJNRPR09}. Therefore, understanding how cooperation evolves is a topic of interest to researchers from diverse fields \cite{axe,LevinER06,OstromScience99,RandTCS13,PercPLA16}.

Over the past decades, evolutionary game theory has been developed \cite{smith} into a powerful approach to study the problem of cooperation in various biological, social and economic settings \cite{TurnerNature99, GintisEHB03, DoebeliEL05, NowakScience04}. It has shown that cooperation flourishes under a variety of mechanisms, including kin selection \cite{HamiltonJTB64}, group selection \cite{TraulsenPNAS06}, and reciprocity \cite{TriversQRB71, NowakNature05, NakamuraPLoSCB11}, just to name a few (we refer to Ref. [21] for a recent review).

Of particular interest is the so-called `network reciprocity' \cite{5rules}, where population structure plays a decisive role in the evolution of cooperation \cite{SzaboPRE98,LiebermanNature05,hetero3,OhtsukiNature06,FloriaPRE09,RocaPRE09,HelbingPNAS09}. Interactions among individuals are not random, but often exquisitely structured, for example, through social networks \cite{JacksonBook08}. This line of research is stimulated largely by the seminal work on spatial games \cite{NowakNature92}, in which individuals are situated on a square lattice and only interact with their immediate neighbours. Ever since the boom of network science \cite{networks1,networks2}, games on networks have been extensively studied (for a comprehensive review we refer to Ref. [33]).

Moreover, population structure itself can also be the consequence of evolution. This notion led to the study of coevolutionary games on networks (or games on dynamical networks), in which both individuals' behavioural strategies and their social connections undergo evolutionary changes (see, for example, Ref. [34] for a recent review). Both theoretical models \cite{Zim,PachecoPRE06,adapt,HanakiMS07,feng1,SkyrmsPNAS09,feng2,SzolnokiEPL09,WuPloS10,DuDGA11,WardilSR14} and behavioural experiments \cite{BsharyBL05,FehlEL11,RandPNAS11,WangPNAS12,ShiradoNC13} show that cooperation can prevail when individuals not only adjust their interaction strategies but also are able to switch their social interaction partners. Furthermore, previous studies reveal that network degree heterogeneity, an important topological feature favouring the evolution of cooperation, can result from partner switching processes \cite{adapt,feng1,DuDGA11}.

When it comes to partner rewiring processes, to the best of our knowledge, little attention has yet been paid to the spatial element of social networks \cite{BarnettPRE07,BarthelemyPhysRep11}. From this perspective, the underlying geographic distance, rather than the graph distance, has an impact on influencing partner choice. Despite vast geographical distance segregating individuals around the globe, the profound small-world phenomenon or six degrees of separation, is ubiquitously found to be a common feature of real-world social networks \cite{watts1,albert2002}. The question then arises: why do some individuals prefer to strike out and form distant partnerships, especially when doing so is potentially costlier than just interacting locally? We will address this question using a coevolutionary model of cooperation and partner rewiring range preference.

Although probed in a different context, prior work has demonstrated the emergence of small world phenomena by randomly rewiring links \cite{watts1} or adding long range links \cite{NewmanPLA99}. In contrast, here we do not rewire networks exogenously by a given parameter, but rather let rewiring be endogenously controlled by individual preference.

Let us now turn to our model. As shown in Figure 1, individuals are situated on a lattice network and engaged in two layers of social interactions that are different in their modifiability. The bottom layer network dictated by lattice distance $l$ is viscous and static, whereas the top layer network comprising of individuals' partners of lattice distance between $l$ and $d$ is fluid, and can be rewired according to their partner rewiring range preferences. 

These layers embody the idea that people form relatively stable strong ties with a core group of close contacts, while having more transient ties with a wider group \cite{Granovetter1973}. In our model, we distinguish these two groups by spatial distance: local vs distant. We work with the simplifying assumptions that rewiring range preferences are either global or local, and that these preferences evolve through social learning and cultural evolution \cite{meso2006, henrich2003}. Our model aims to capture reasonable features of a social system, which we will use to find potential explanatory factors for the creation of connections between geographically distant individuals.

We give individuals agency using an existing evolutionary game theoretical framework \cite{feng1,feng2}. Under this framework, individuals play the prisoner's dilemma \cite{PD} with their partners, choosing to cooperate (C) or defect (D). When two individuals both cooperate, they each receive a payoff of $1$. If they both defect, each of them instead receive a payoff of $u$, where $0 < u < 1$. Finally, if one cooperates and the other defects, the cooperator gets $0$ while the defector gets $1 + u$. The parameter $u$ represents the cost-benefit ratio of cooperation, and would generally result in a greater proportion of defectors when increased \cite{feng1,feng2,adapt,5rules}.

At each discrete time step, a random individual, with probability $w$, decides to rewire a connection from a low reputation partner to a new partner with potentially higher reputation, found within her rewiring range preference (global or local). Local rewiring incurs no cost, but global rewiring incurs a one-time cost $c$. The parameter $p \in [0,1]$ encodes the probability that she is able to find the high reputation individual and create a new link. While with probability $1-p$, she is unable to do so and simply creates a new link to a random individual within her rewiring range preference.

Otherwise, with probability $1-w$, the individual compares her payoff with that of a neighbour, via the Fermi equation \cite{update1,update2,update3}, and decides whether to copy both the neighbour's strategy and rewiring range preference. In general, larger values of $w$ allow links to low reputation defectors to be severed rapidly and hence promote cooperation \cite{feng1,feng2,adapt}. 

The model, as well as our simulations, are described in greater details in the methods section.

\section{Results}

First, in Figure 2, we compare the cases where individuals have Figure 2a only global rewiring range preference, Figure 2e only local rewiring range preference, and Figure 2i when we have coevolution of strategy, C or D, with rewiring range preference, local or global. In Figure 2i, individuals are initially assigned local or global rewiring range preference with equal probability.

Figure 2a shows that when individuals have only global rewiring range preference, cooperation arises. Snapshots Figure 2 b,c,d of the log number of partners for each individual taken at different point in time, shows that global rewiring helps cooperators attract partners across the entire population and grow into hubs. This promotes degree heterogeneity (see Supplementary Figure 1), and hence cooperation \cite{hetero1,hetero2,hetero3}. On the other hand, Figure 2e shows that cooperation is not favoured when individuals are restricted to only having local rewiring range preferences. As shown in Figure 2g, the ability to rewire from low reputation partners to ones with potentially higher reputation aids cooperation in the short run. As $d = 2$ and $l =1$, when node range preferences are local, new partners must be within distance $3$. Consequently, the maximum number of partners under these assumptions is $24$. This puts a limit on the level of heterogeneity (see Supplementary Figure 1), makes the population more vulnerable to invasion by defectors, and is deleterious to cooperation in the long run \cite{hetero1,hetero2,hetero3}.

When we have coevolution of strategy with rewiring range preference, Figure 2i shows that heterogeneity emerges in the short run and continues to develop in the long run, allowing cooperators to attract long range partners, and be dominant at equilibrium. Hence, we see that just having the option of long range partner rewiring favours cooperation in the long run. In Figure 3, we see similar results for a wide variety of parameters combinations.

Supplementary Figure 2 compares the time evolution of the average path length and the average clustering coefficient \cite{watts1} of these three cases with that of an Erd\H{o}s-R\'{e}nyi (ER) random graph \cite{erdos} ensemble with the same average degree. Supplementary Figure 11 shows how the average path length varies with number of individuals $n$. When rewiring range preference is only global or allowed to coevolve with strategy, the average distance between individuals in the resulting network is close to that of the ER random graph ensemble, while the average clustering coefficient is close to an order of magnitude higher, and the average distance between individuals grows at a rate proportional to $\log{n}$. Consequently, we classify these as small-world networks. On the other hand, with only local rewiring preferences, the average distance between individuals in the resulting network fails to be close to that of the ER random graph ensemble, grows faster than $\log{n}$, and hence does not have the small-world property.

Figure 3a provide examples of combinations of $u$ and $w$ which favour cooperators with global range preferences. When this happens, we have an abundance of long range links, which is often sufficient for the small-world phenomenon to emerge \cite{watts1}. In particular, these global cooperators thrive under fast partner switching $w$ and small cost-benefit ratio $u$, as shown by the lighter blue patches in the top left corner of Figure 3a. When the probability of switching, $w$, is high, global cooperators rapidly delete links to defectors, and reliably find or attract other cooperators across great distances. A small cost-benefit ratio, $u$, makes it less likely for cooperators to become defectors, giving an opportunity for global cooperators' advantage over local cooperators to take effect.

On the other hand, the orange patches in the bottom right corner shows that global defectors are dominant in the case of slow partner switching $w$ and high cost-benefit ratio $u$. When $w$ is low, cooperators cannot rapidly delete links to defectors. Furthermore, a high $u$ influences the partners of defectors to turn into defectors. This saturates their local neighbourhoods with defectors, which is advantageous to global defectors who are able to seek out cooperators across the entire network. In these cases, even though long range links are plentiful, the inability to sustain cooperation make these scenarios undesirable as models of real world social systems.

In the parameter space where our evolutionary dynamics transits from most favouring global cooperators to global defectors, local cooperators can be most favoured. Although cooperation evolves under these conditions, the higher cost-benefit ratio $u$ combined with the existence of a global rewiring cost $c$ penalizes global cooperators in favour of local cooperators. From Figure 3b and Figure 4b, we can see that even though global cooperators are necessary for the evolution of cooperation, they eventually switched to having a local rewiring range preference, because global cooperators could not form enough cooperative partnerships to offset the accumulated cost of global rewiring $C$, especially when the higher $u$ encourages the population to have a larger proportion of defectors.

We also observe that even when defection is suppressed, global defectors are often more abundant than local defectors, due to their ability to find high reputation cooperators across the entire network. Examples of this are seen in Figure 3b and c, which illustrates how the proportion of the four types of individual, global vs local and cooperator vs defector, evolves over time.

In Figure 4a, we extend our analysis of favourable conditions for global cooperators by varying parameter $p$, while parameters $u = 0.2$ and $w = 0.2$ are fixed. In general, we see cooperation rises with $p$, as cooperators can more reliably find other cooperators after deleting a low reputation link. In particular, this increased availability of reputation information helps global cooperators form clusters of cooperators (see Supplementary Figure 3) through reputation based rewiring \cite{CoopClust}. Despite that, the equilibrium frequency of global cooperators are maximized at an intermediate availability of reputation information, $p = 0.15$, illustrated in Figure 4b.

As availability of reputation information increases beyond $p = 0.15$, we see a lower proportion of defectors in the population in general, as cooperators are reliably able to sever low reputation links and connect to higher reputation individuals. This, together with the abundance of reputation information, allows local cooperators to more reliably rewire to cooperators despite not having the ability to find partners globally. The combination of these factors and the cost of global rewiring contribute to the decline of global cooperators in favour of local cooperators. This narrative is seen most clearly for large values of $p$, for example in Figure 4c $p = 0.5$, in which global cooperators dominate the population initially, driving defectors to extinction, but is soon overtaken by local cooperators due to the cost of global rewiring.

Hence, the goal of maximizing the proportion of global cooperation and long range links can be accomplished only when there is intermediate availability of reputation information, and not when such information is lacking or in abundance.

\section{Discussion}

In this paper, we proposed a mechanism for the emergence of the small world property, based on evolutionary dynamics, with assumptions that are plausible in real world networks. We do so by first placing individuals on a spatial lattice. We then use the lattice distance to define rewiring range preferences for each individual. In this setting, we studied the coevolution of strategy and rewiring range preference, in the presence of long range rewiring costs. 

We find that despite the presence of costs, global cooperators can be favoured for some parameter combinations. For other parameter combinations, local cooperators can be favoured. In particular, for the parameters in Figure 3a, given any cost-benefit ratio $u$, as the rewiring probability $w$ is increased, there is a point where local cooperators first become dominant followed by global cooperators.

We focused on the more realistic scenarios in which cooperation is favoured at equilibrium, and observed that these cooperators with global rewiring preferences cause degree heterogeneity to emerge in the resulting networks. Some of these individuals become hubs, with many long distance links, contributing to the small world property. We also found that cooperators with global rewiring preference are maximized at intermediate availability of reputation information, and are discouraged at low or high levels of $p$.

In our model, the cost of rewiring can be interpreted as the cost of searching globally for a new partner, paid for by only the proposer. The recipient always accepts new partners because there are no costs, and new links are always potentially beneficial under our parameterisation of the prisoner's dilemma. In the supplementary, we also looked at results from one possible way to model bilateral link creation, even though the convention is for link creation to be unilateral \cite{SzaboPhysRep07, perc, feng1}. In the bilateral link creation version of our model, the recipient only accepts a link if the proposer is within her rewiring range preference, and both parties pay a link creation cost $c$ if the distance between them is greater than local rewiring, $d+1$. 

We obtain the same results seen in Figure 2 (see Supplementary Figure 12). For bilateral link creation, there is also a region in the $(u,w)$ parameter space where coevolutionary dynamics transits from most favouring global cooperators to global defectors. However, local range preferences are never favoured, and global range preference dominates for all parameter combinations (see Supplementary Figure 13). Proportion of global cooperators increases with $p$, and local cooperators are suppressed (see Supplementary Figure 14). When link creation is bilateral, those with local rewiring range preferences are unable to compete because they lose the ability to receive long range links, while those with global preferences gain the ability to rewire locally without cost.

A natural next step for future investigation is to consider more realistic coevolutionary models that assign initial rewiring range preferences uniformly at random from a range of discrete values, and rewiring costs to be a possibly non-linear function of range preferences. Preliminary simulations point to conditions that encourages cooperators with rewiring range preferences greater than the average of the initial assignment (see Supplementary Figure 7). We hope that further study of this more general model would not only lead to more comprehensive results on the coevolution of small world and cooperation phenomena, but the spatial setting could potentially allow us to study the emergence of community structures that are observed in real world networks. \cite{comm1,comm2}.

\section{Methods}

\subsection{Model \\}

The $n = m^2$ individuals in our model are situated on a two dimensional $m \times m$ square lattice with periodic boundaries. The lattice distance between two nodes $x$ and $y$, with coordinates $(a,b)$ and $(c,d)$ respectively, is the length of the shortest path between them, $d(x,y) = |a-c| + |b-d|$. 

We derive a network from this setup by connecting individuals within lattice distance $d$. Links to individuals within distance $l$ are local static links, which cannot be changed or rewired. Global links, which are not static and can change via the evolutionary dynamics, join individuals with distances between $l$ and $d$.

Individuals interact by playing the Prisoner's Dilemma \cite{PD} with their partners. They are initially assigned to be a cooperator $C = [1,0]^T$ or a defector $D = [0,1]^T$ with equal probability. The payoff matrix for our version of prisoner's dilemma is,

$$\bordermatrix{~ & C & D \cr
                  C & b-c & -c \cr
                  D & b & 0 \cr},$$

where $b,c$ are parameters representing the respective benefit and cost of cooperation. We can re-write the matrix with a single parameter \cite{param1,param2} as,

$$M = \bordermatrix{~ & C & D \cr
                  C & 1 & 0 \cr
                  D & 1+u & u \cr},$$

where $u \in [0,1]$ is the cost-benefit ratio of cooperation. Let $\mathcal{N}_i$ be the neighbourhood of individual $i$ on the graph, then the total payoff of individual $i$ is,

$$ P_i = \sum_{j \in \mathcal{N}_i} s_i^T M s_j,$$

where $s_i$ is the strategy, $C$ or $D$, of individual $i$.

The reputation $R_i(t)$ of an individual $i$ at a discrete time step $t$ is the number of times she has cooperated prior to $t$. This is defined as, 

$$R_i(t+1) = R_i(t) + \delta_i(t),$$

where $\delta_i(t) = 1$ if the individual $i$ cooperates at the end of the $t$-th time step, and zero otherwise. At each time step, we pick an individual uniformly at random. With probability $w$, she chooses to switch one of her partners, otherwise with probability $1-w$, she updates her strategy.

{\bf Partner switching}: When individual $i$ chooses to switch partners, she attempts to maximize the gain from the switch by removing a partner with low reputation, and adding one with higher reputation. 

Individuals are initially assigned to have local or global rewiring range preference with equal probability. While local rewiring incurs no cost, global rewiring incurs a one-time cost $c$ to model the cost of collecting reputation information. Every time an individual $i$ with global rewiring preference switches partner, this cost is incurred and a variable $C_i(t)$ is incremented by $c$ until the next time she takes part in a strategy update comparison.

As the individual is forbidden from severing links with partners within distance $l$, she first isolates an existing partner more than distance $l$ away who has the least reputation. She then severs that connection, or in the case of a tie, severs a connection at random from among the tied individuals.

Next, she searches for an individual that she is not connected with, within her rewiring range preference, who has the largest reputation. A parameter $p \in [0,1]$ encodes how easily the individual can access and search reputation information. With probability $p$, she is able to find the high reputation individual and create a new link. While with probability $1-p$, she is unable to do so and simply creates a new link to a random individual within her rewiring range preference.

{\bf Strategy updating}: When individual $i$ chooses to update her strategy, she chooses a partner $j$ uniformly at random and decides whether to emulate the partner's strategy by comparing their total payoffs, after both party have subtracted the one-time costs $c$ they have accumulated up to this point. Both of $j$'s strategy and rewiring range preference replaces those of individual $i$ with probability given by the Fermi function \cite{update1,update2,update3},

$$ \phi(s_i,r_i \leftarrow s_j,r_j) = \frac{1}{1+\exp[\beta \{ (P_i - C_i(t)) - (P_j - C_j(t)) \} ]}.$$

where $\beta$ is a parameter such that $0 \leq \beta$, and $r_i$ is the rewiring range preference, local or global, of individual $i$. Regardless of whether individual $i$ emulates her partner's strategy and rewiring range preference, both $C_i$ and $C_j$ are set to zero because they represent temporary search and link creation costs.

\subsection{Simulation \\}

Each row in Figure 2 is taken from a single run. In Figure 2 a,b,c,d, we modify our model to allow individuals to only have global rewiring range preference. For Figure 2 e,f,g,h, we do the same and allow individuals to only have local rewiring range preference. In Figure 2 i,j,k,l, individuals have initial rewiring range preference local or global with equal probability. Parameters for Figure 2 are $n = 400$, $t = 1.2 \times 10^5$, $\beta = 0.1$, $u = 0.1$, $w = 0.5$, $p = 0.1$, $c = 0.2$, $d = 2$ and $l =1$.

In Figure 3a, each combination $(u,w)$ in the parameter space is run for $2 \times 10^6$ time steps, and the results are averaged over an additional $3 \times 10^4$ time steps. This process is then repeated for a total of $100$ runs, and the results are averaged again over these $100$ runs. The colours for each point in Figure 3a are determined by first determining if cooperation or defection is in the majority ($\geq 50 \%$). Then, we calculate the fraction of individuals in this majority that have global rewiring range preference, and assign the colours based on the scale in Figure 3a. For instance, dark blue illustrates the dominance of local cooperators in the majority, while light orange illustrates the dominance of global defectors in the majority. Figure 3 b,c are taken from the simulations in Figure 3a. Other parameters for Figure 3 are $n = 3600$, $\beta = 0.1$, $p = 0.1$, $c = 0.2$, $d = 2$ and $l =1$.

In Figure 4a, each value of $p$ is run for $2 \times 10^6$ time steps, and the results are averaged over an additional $3 \times 10^4$ time steps. This process is then repeated for a total of $100$ runs, and the results are averaged again over these $100$ runs. Figure 4 b,c are taken from the simulations in Figure 4a. Other parameters for Figure 4 are $n = 3600$, $u = 0.2$, $w = 0.2$, $\beta = 0.1$, $c = 0.2$, $d = 2$ and $l =1$.

Various values for $c$, the cost of long range rewiring, have been considered (see Supplementary Figure 8, 9 and 10), and $c = 0.2$ was chosen as representative of the general trends.

\clearpage

\section*{Acknowledgements} 

This work was supported by Dartmouth Faculty Start-up Fund to F.F. We thank the reviewers for their comments an earlier version of the manuscript, which helped us improve our work. \\

[Author Contributions] T.K. and F.F. conceived the model, T.K. wrote the code, conducted the experiments, and analysed the results. All authors reviewed the manuscript. \\

[Competing Interests] The authors declare that they have no competing financial interests. \\

\clearpage

\textbf{Figure captions:} \small
\begin{description}
  \item[Figure~1]{\bf Model schematic.} Individuals are situated on a square lattice with periodic boundaries, and are engaged in two layers of interactions. We consider two traits: game strategies, cooperate vs defect, and partner rewiring range preferences, local vs global. Global rewiring has no distance restriction but incurs a cost $c$. Local rewiring is restricted to lattice distance $d+1$. In the lower layer, interactions are within lattice distance of one, are static, and cannot be rewired. In the upper layer, interactions are more than distance one away, and can be rewired based on reputation. Cooperators tend to have more partners than defectors in the upper layer, and therefore can be dominant in the population. 

 \item[Figure~2] {\bf Comparison of global rewiring, local rewiring and both under coevolution.} The first column shows the fraction of cooperators over the course of evolution with Fig 2a only global rewiring, Fig 2e only local rewiring, and i) both under coevolution. Having the option of long range partner rewiring in Fig 2a and Fig 2i, even at a cost $c$, makes it easier for cooperation (blue) to evolve, than in Fig 2e. Snapshots Fig 2 b,c,d taken from Fig 2a at different time points shows that global rewiring helps cooperators attract partners across the entire population, and grow into hubs. In contrast, local rewiring Fig 2 f,g,h helps cooperation in the short run but the lack of heterogeneity (Supplementary Figure 1) makes it more vulnerable to invasion by defectors (orange). When partner rewiring preference is an evolvable trait, Fig 2 j,k,l, under certain conditions, global cooperators will be favoured by natural selection and emerge as skyscrapers. Supplementary Figures 2 and 3 shows the resulting networks for only global rewiring and coevolution having small average shortest path length while maintaining high clustering, thereby possessing the small-world properties. Parameters: $n=400$, $t = 1.2 \times 10^5$, $\beta=0.1$, $u = 0.1$, $w = 0.5$, $p=0.1$, $c=0.2$, $d=2$ and $l=1$.

\item[Figure~3]{\bf Coevolution of cooperation and long range partner rewiring preferences.} Fig 3a shows the most common traits across the parameter space $(u,w)$. Global cooperators are most favoured under fast partner switching $w$ and small cost-benefit ratio $u$. Whereas global defectors are dominant for low $w$ and high $u$. In the parameter space where coevolutionary dynamics transits from most favouring global cooperators to global defectors, local cooperators can be most favoured. Fig 3b and Fig 3c shows how the proportion of the four types of individual evolves over time for the two pairs of parameters ($\Box$) chosen in Fig 3a. The most favoured behaviour type is local cooperation in Fig 3b vs global cooperation in Fig 3c. Even when defection is suppressed, global defectors are more abundant than local defectors. Parameters: $n = 3600$, $t = 2 \times 10^6$, $\beta = 0.1$, $p = 0.1$, $c = 0.2$, $d =2$ and $l=1$.

\item[Figure~4] {\bf Optimum availability of reputation information for the evolution of global cooperation.} Fig 4a shows the proportion of local and global cooperators at equilibrium as a function of availability of information, $p$. An increase in $p$ always helps cooperation. Yet, intermediate availability of reputation information maximizes the equilibrium frequency of global cooperators. Fig 4 b,c shows the coevolutionary dynamics of the four types of individuals over time, for Fig 4b intermediate $p$, and Fig 4c large $p$. Global cooperators benefit from an increase in availability of reputation information as it helps them form clusters of cooperators through reputation based rewiring. At large values of $p$, global cooperators dominate the population initially, and drive defectors to extinction, but is soon overtaken by local cooperators due to the cost of global rewiring. In all cases, global cooperators are the catalyst for cooperation. Parameters: $n = 3600$, $t = 2 \times 10^6$, $u = 0.2$, $w = 0.2$, $\beta = 0.1$, $c = 0.2$, $d =2$ and $l=1$.

\end{description}

\newpage
\textbf{Supporting information figure captions:} \small
\begin{description}

  \item[Figure~S1] {\bf Degree heterogeneity over time.} Figures show the variance of the degree distribution over time for only global rewiring, only local rewiring, and both under coevolution. In both Sup. Fig. 1 a and b, networks produced by only global rewiring and coevolution have degree distributions that increase in variance, and hence degree heterogeneity, over time. In contrast, networks produced by only local rewiring do not achieve the same level of variance. The network produced by coevolution in Sup. Fig. 1 b has a variance that eventually outgrow that of the network produced by only global rewiring. Simulation data in Sup. Fig. 1a was also used for Figure 2 in the main text. Parameters: $u=0.1$, $w = 0.5$, $\beta = 0.1$, $p = 0.1$, $c = 0.2$, $d = 2$, $l = 1$, single run.
  
   \item[Figure~S2] {\bf Comparison of average node distance.} Figures show the time evolution of average node distance (average of the shortest distance between nodes) for networks produced by only global rewiring, only local rewiring, and both under coevolution. In both Sup. Fig. 2 a and b, networks produced by only global rewiring and coevolution have average node distance close to that of an Erd\H{o}s-R\'{e}nyi (ER) random graph ensemble with the same average degree. On the other hand, average node distance remains much higher with only local rewiring. This is difference is more pronounced as the network size is increased from Sup. Fig. 2a $n = 400$ to Sup. Fig. 2b $n = 1600$. Simulation data in Sup. Fig. 2a was also used for Figure 2 in the main text. Parameters: $u=0.1$, $w = 0.5$, $\beta = 0.1$, $p = 0.1$, $c = 0.2$, $d = 2$, $l = 1$, single run.
   
   \item[Figure~S3] {\bf Comparison of average clustering coefficient.}  Figures show the average clustering coefficient for networks produced by only global rewiring, only local rewiring, and both under coevolution. In Sup. Fig. 3b, networks produced by only global rewiring and coevolution have average clustering coefficient at least one order of magnitude greater than that of an Erd\H{o}s-R\'{e}nyi (ER) random graph ensemble with the same average degree. In Sup. Fig. 3a, these statistics are close to one order of magnitude greater. With only local rewiring, the clustering coefficient remains many orders of magnitude greater than the ER random graph ensemble in both Sup. Fig. 3a and Sup. Fig. 3b. Simulation data in Sup. Fig. 3a was also used for Figure 2 in the main text. Parameters: $u=0.1$, $w = 0.5$, $\beta = 0.1$, $p = 0.1$, $c = 0.2$, $d = 2$, $l = 1$, single run.
    
   \item[Figure~S4] {\bf Strategy and density of highest degree nodes.} Sup. Fig. 4a shows the top $10 \%$ highest degree nodes becoming all cooperators early in the simulation. Sup. Fig. 4a is truncated at $t = 12 \times 10^3$ as the fraction of cooperators remained close to $1$ for the rest of the simulation. Sup. Fig. 4b shows the ratio of the edge density (number of edges divided by number of possible edges) of the top $10 \%$ highest degree nodes to that of the entire network. Sup. Fig. 4b is truncated at $t = 15 \times 10^5$ as the ratio remained close to $12$ for the rest of the simulation. Parameters:  $n = 3600$, $t = 2 \times 10^6$, $u=0.2$, $w = 0.2$, $\beta = 0.1$, $p = 0.15$, $c = 0.2$, $d = 2$, $l = 1$, single run.

  \item[Figure~S5] {\bf Optimal  availability of reputation information for the evolution of global defectors.} A situation similar to Figure 4 happens for defectors. The proportion of global defectors increases with $p$, and reaches a global maximum, before falling again. Simulation data was also used for Figure 4 in the main text. Parameters: $n = 3600$, $t = 2 \times 10^6$, $u=0.2$, $w = 0.2$, $\beta = 0.1$, $c = 0.2$, $d = 2$, $l = 1$, final result averaged over an additional $3 \times 10^4$ time steps, this process is then repeated for 100 runs.

 \item[Figure~S6] {\bf Global vs local cooperators at high values of $p$.} As parameter $p$ is increased beyond $p = 0.5$, proportion of global cooperators increases monotonically. Simulation data was also used for Figure 4 in the main text. Parameters: $n = 3600$, $t = 2 \times 10^6$, $u=0.2$, $w = 0.2$, $\beta = 0.1$, $c = 0.2$, $d = 2$, $l = 1$, final result averaged over an additional $3 \times 10^4$ time steps, this process is then repeated for 100 runs.

 \item[Figure~S7] {\bf Non-binary rewiring range preference.} In Sup. Fig. 7 a and b, rewiring range preference is allowed to take values in $[d+1,n]$, initially assigned uniformly at random. The cost of rewiring is instead $(\frac{r_i}{n})^\alpha$, where $r_i$ is the rewiring range preference of individual $i$, and Sup. Fig. 7a $\alpha = 0.5$ and Sup. Fig. 7b $\alpha = 1$. In both cases, the distribution of equilibrium rewiring range preference have a median, $18$ and $19$, that is higher than the mean and median, $16$, of the uniform distribution on $\{3,4,\dots,29,30\}$. Parameters: $n = 900$, $t = 2 \times 10^6$, $u=0.3$, $w = 0.8$, $\beta = 0.1$, $p = 0.1$, $c = 0.2$, $d = 2$, $l = 1$, combined over $100$ runs.

 \item[Figure~S8] {\bf Global rewiring (green), local rewiring (red) and both under coevolution (blue), with different costs $c$.} Parameters: $n = 3600$, $t = 1 \times 10^6$, $u=0.1$, $\beta = 0.1$, $p = 0.1$, $d = 2$, $l = 1$, averaged over $20$ runs, lattice with non-periodic boundaries.

 \item[Figure~S9] {\bf Time evolution of proportion of GC, LC, GD, and LD under $c = 0.1$.} Parameters: $n = 400$, $t = 5 \times 10^4$, $\beta = 0.01$, $p = 0.1$, $c = 0.1$, $d = 2$, $l = 1$, averaged over $100$ runs, lattice with non-periodic boundaries.

 \item[Figure~S10] {\bf Time evolution of proportion of GC, LC, GD, and LD under $c = 0.5$.} Parameters: $n = 400$, $t = 5 \times 10^4$, $\beta = 0.01$, $p = 0.1$, $c = 0.5$, $d = 2$, $l = 1$, averaged over $100$ runs, lattice with non-periodic boundaries.

 \item[Figure~S11] {\bf Change in average path length as number of nodes increases.} Figures show the average path length of the network at equilibrium as the number of nodes $n$ increases, for only global rewiring, only local rewiring, and both under coevolution. Sup. Fig. 11a shows the results for the unilateral link creation case, while Sup. Fig. 11b shows the results for the bilateral link creation case. In both Sup. Fig. 11a and Sup. Fig. 11b, when we have only global rewiring or both under coevolution, average path length grows at a rate that is lower than $\log{n}$. When we have only local rewiring, the average path length grow at a rate higher than $\log{n}$. Parameters: $u = 0.1$, $w = 0.5$, $\beta=0.1$, $p=0.1$, $c=0.2$, $d=2$ and $l=1$, each averaged over at least $5$ runs.

 \item[Figure~S12] {\bf Bilateral link creation comparison of global rewiring, local rewiring and both under coevolution.} Figure shows the same comparison done in Figure 2 in the main text, but with the bilateral link creation instead. The results are the same as those for the case of unilateral link creation. Parameters: $n=400$, $t = 1.2 \times 10^5$, $\beta=0.1$, $u = 0.1$, $w = 0.5$, $p=0.1$, $c=0.2$, $d=2$ and $l=1$.

 \item[Figure~S13] {\bf Coevolution of cooperation and long range partner rewiring preferences under bilateral link creation.} Figure shows the most common traits across the parameter space $(u,w)$. This is same as the comparison done in Figure 3 in the main text, but for the case of bilateral link creation instead. Global cooperators are most favoured under fast partner switching $w$ and small cost-benefit ratio $u$. Whereas global defectors are dominant for low $w$ and high $u$. Like for Figure 3, there is a region in the parameter space where coevolutionary dynamics transits from most favouring global cooperators to global defectors. However, local cooperators can no longer be favoured. Parameters: $n = 2500$, $t = 1.5 \times 10^6$, $\beta = 0.1$, $p = 0.1$, $c = 0.2$, $d =2$ and $l=1$, end results averaged over an additional $3 \times 10^4$ time steps for each run, total $50$ runs.

 \item[Figure~S14] {\bf Availability of reputation information and the evolution of global cooperation under bilateral link creation.} Figure shows the proportion of local and global cooperators, at equilibrium, as a function of availability of information, $p$, in the case of the bilateral tie creation model. In general, the proportion of global cooperators increase with $p$, while local cooperators are never favoured. Parameters: $n = 2500$, $t = 1.5 \times 10^6$, $u = 0.2$, $w = 0.2$, $\beta = 0.1$, $c = 0.2$, $d =2$ and $l=1$, end results averaged over an additional $3 \times 10^4$ time steps for each run, total $50$ runs.

\end{description}

\newpage

\begin{figure}[!ht]
\begin{center}
\includegraphics[width = \columnwidth]{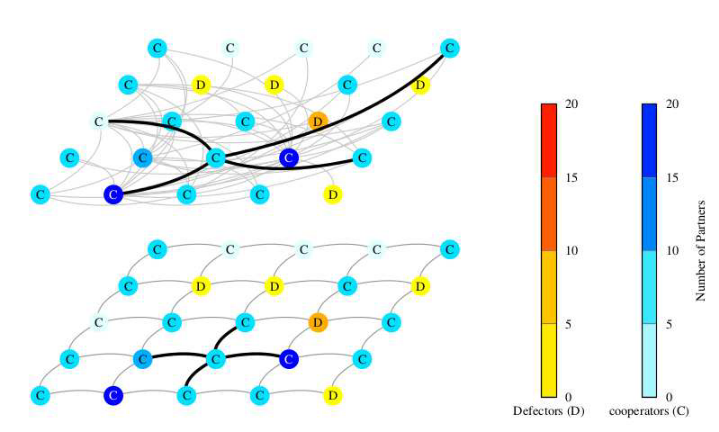}
\end{center}
\caption{}
\label{scheme}
\end{figure}

\begin{figure}[!ht]
\begin{center}
\includegraphics[width = \columnwidth]{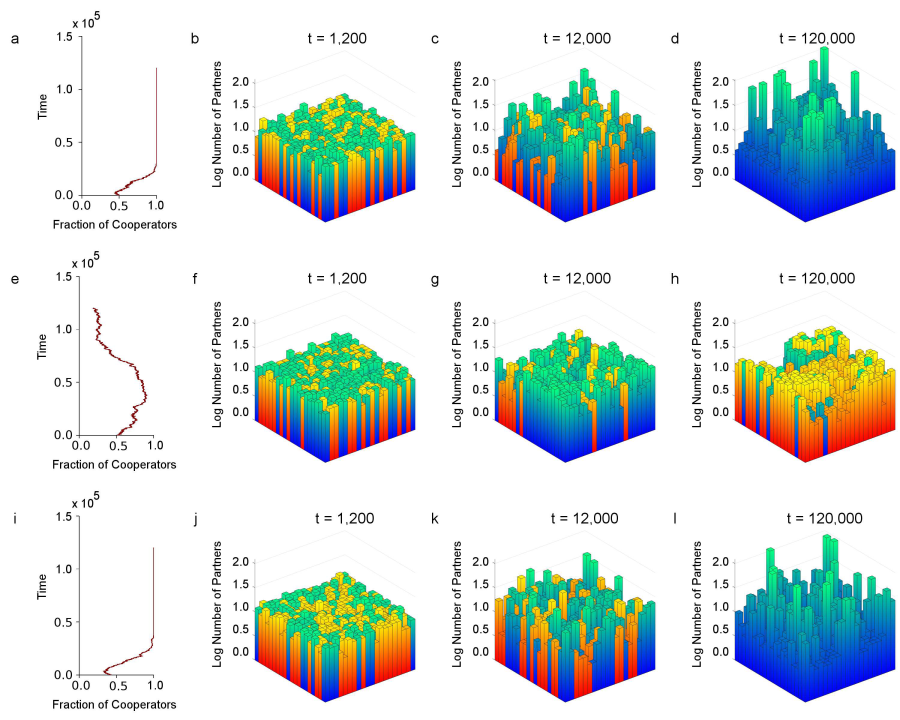}
\end{center}
\caption{}
\end{figure}

\begin{figure}[!ht]
\begin{center}
\includegraphics[width = \columnwidth]{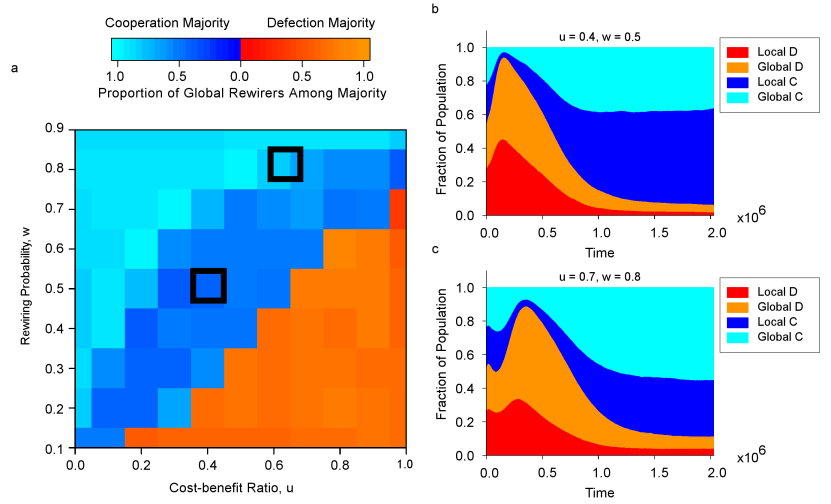}
\end{center}
\caption{}
\end{figure}
\clearpage

\newpage
\begin{figure}[!ht]
\begin{center}
\includegraphics[width = \columnwidth]{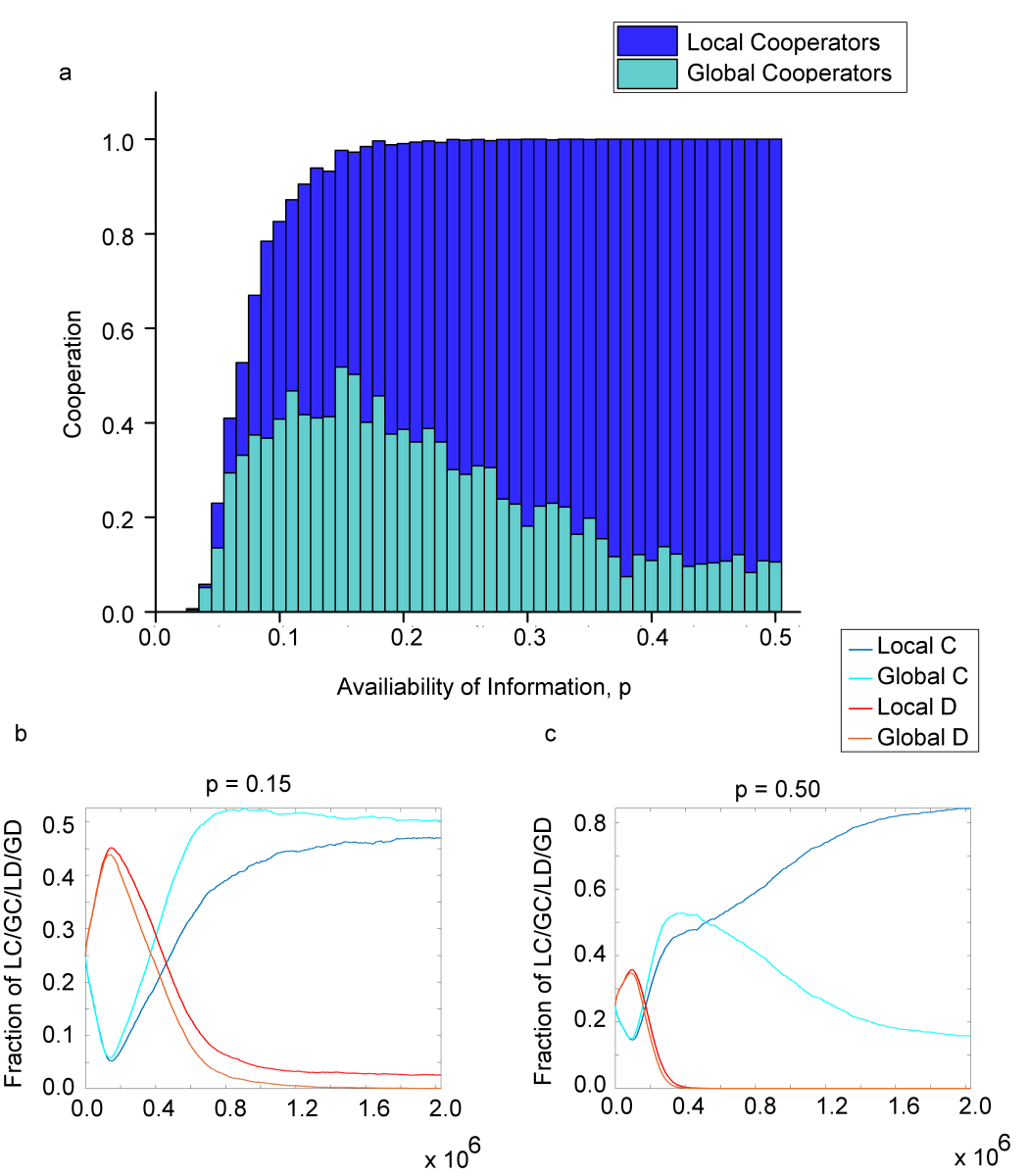}
\end{center}
\caption{
}
\end{figure}

\newpage
\renewcommand{\figurename}{FigS.}
\setcounter{figure}{0} 

\begin{figure}[!ht]
\begin{center}
\includegraphics[scale=0.4]{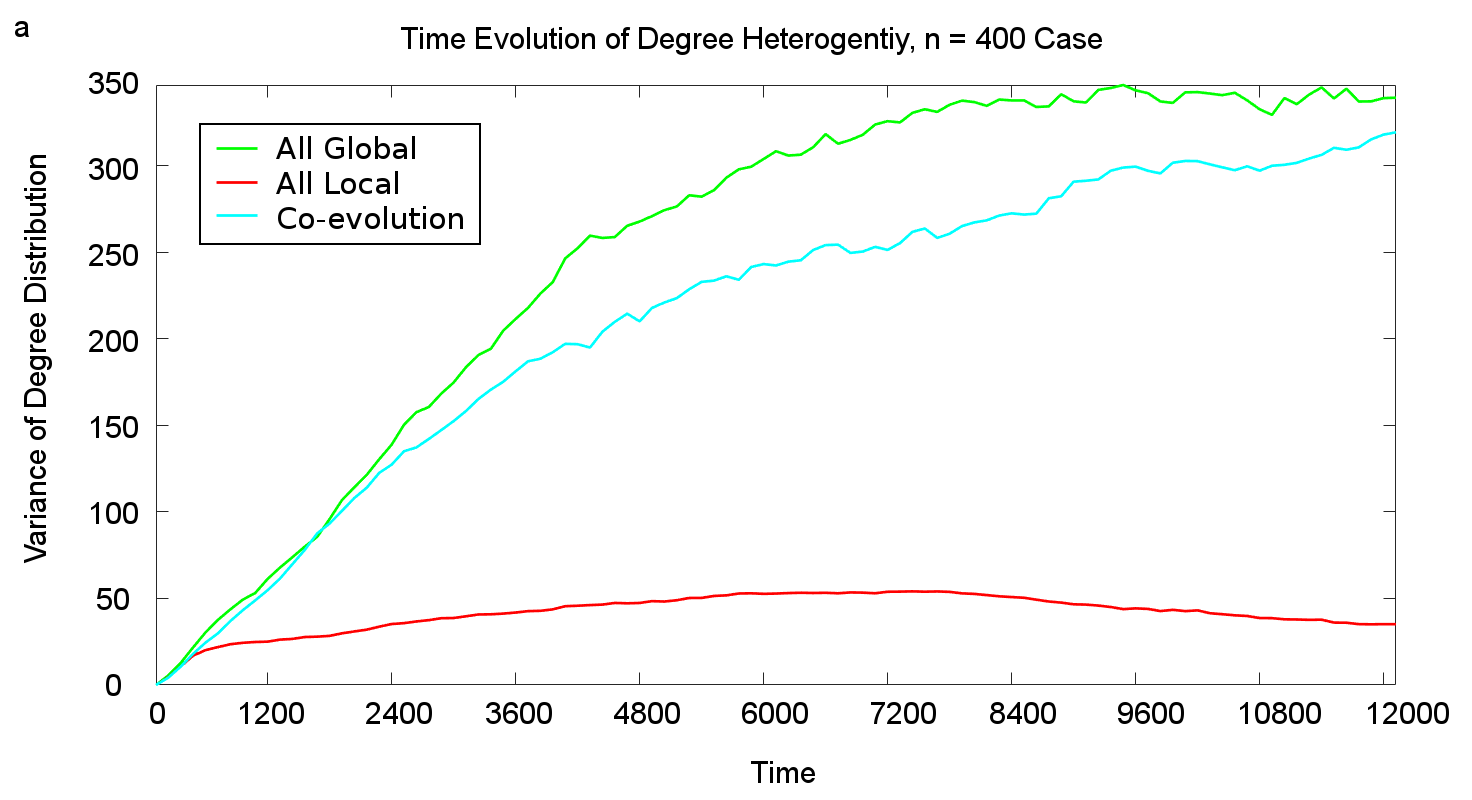}
\includegraphics[scale=0.4]{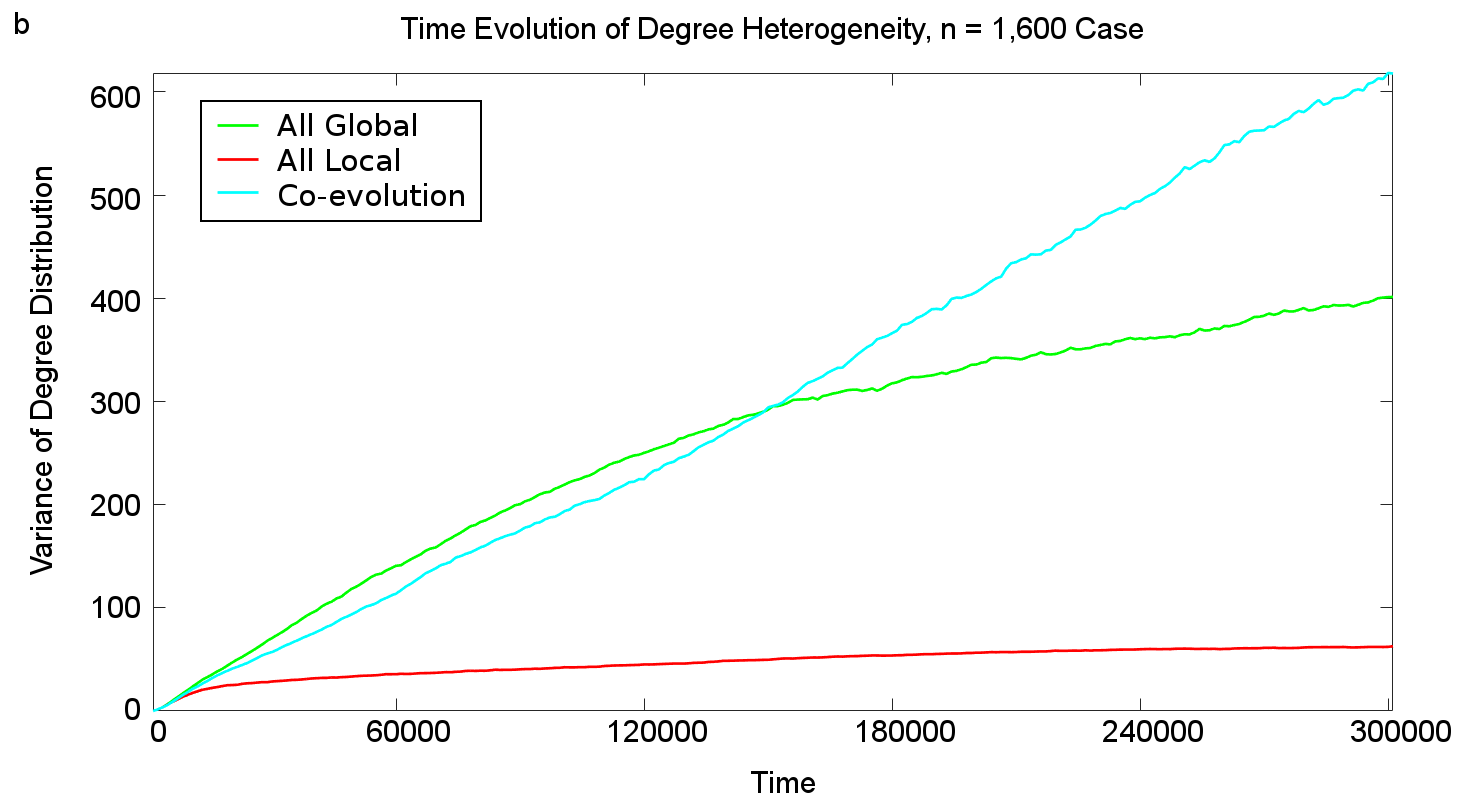}
\end{center}
\caption{}
\end{figure}

\begin{figure}[!ht]
\begin{center}
\includegraphics[scale=0.42]{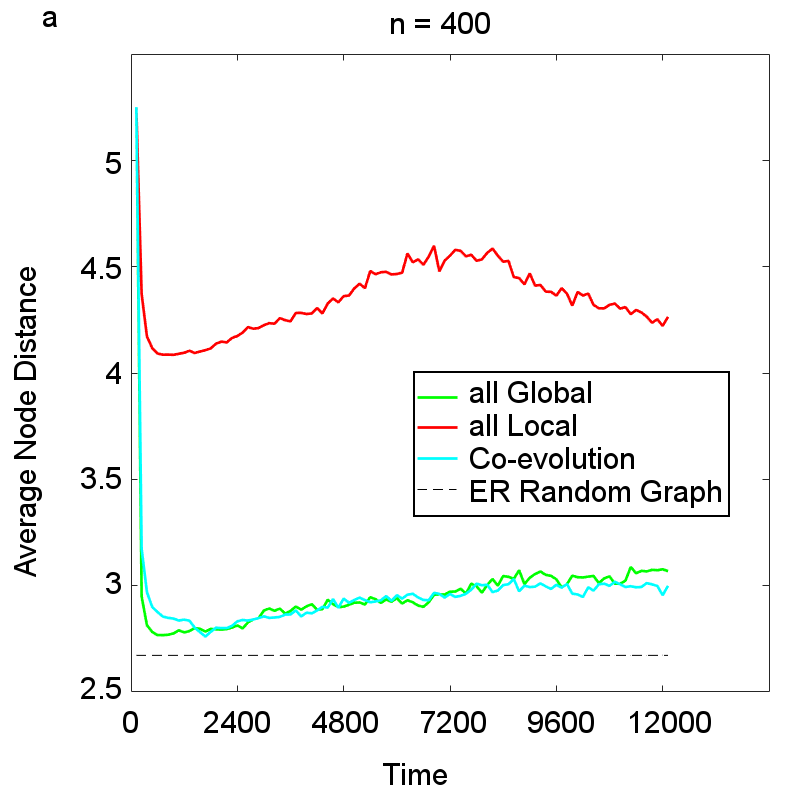}
\includegraphics[scale=0.42]{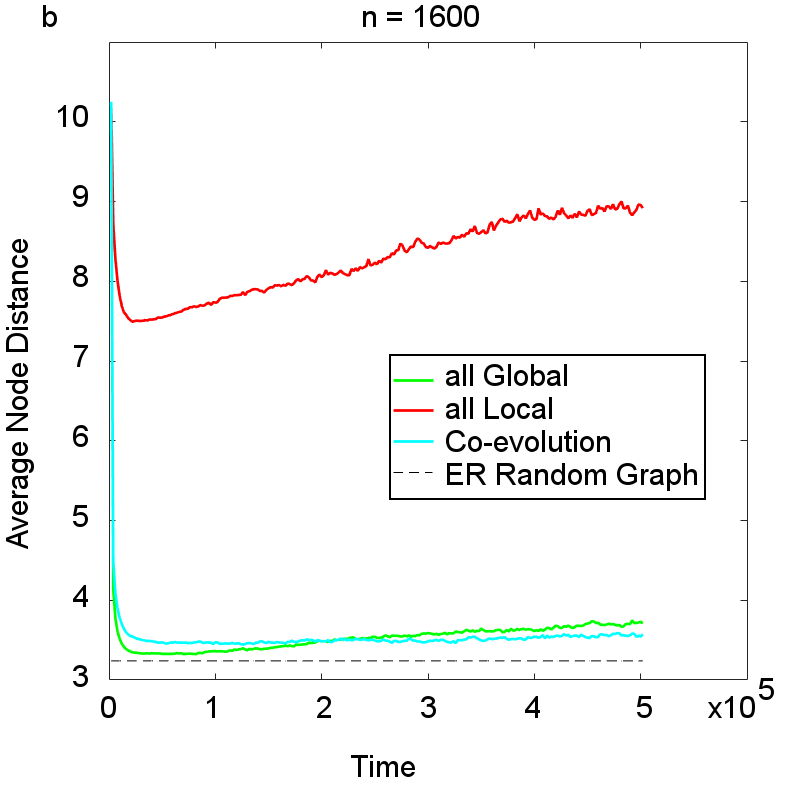}
\end{center}
\caption{}

\end{figure}
\begin{figure}[!ht]
\begin{center}
\includegraphics[scale=0.42]{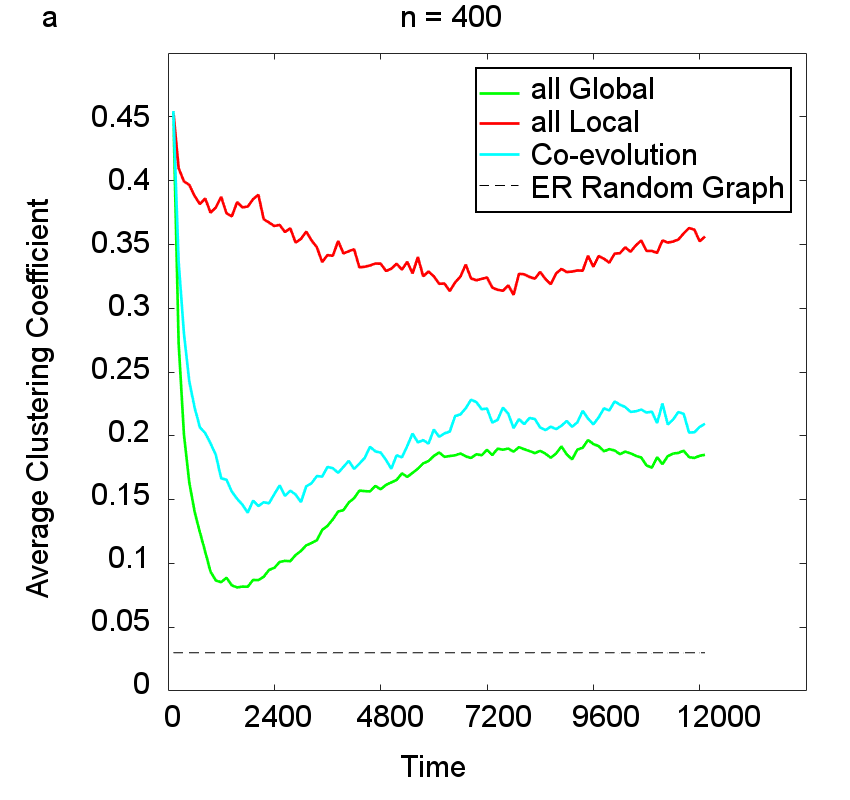}
\includegraphics[scale=0.42]{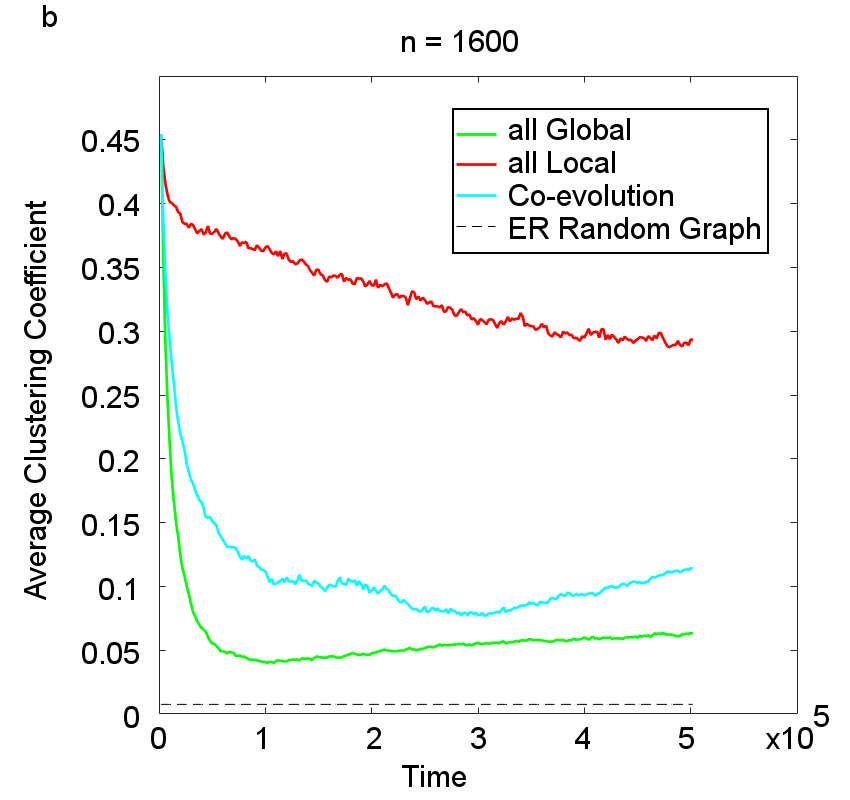}
\end{center}
\caption{}
\end{figure}

\begin{figure}[!ht]
\begin{center}
\includegraphics[scale=0.45]{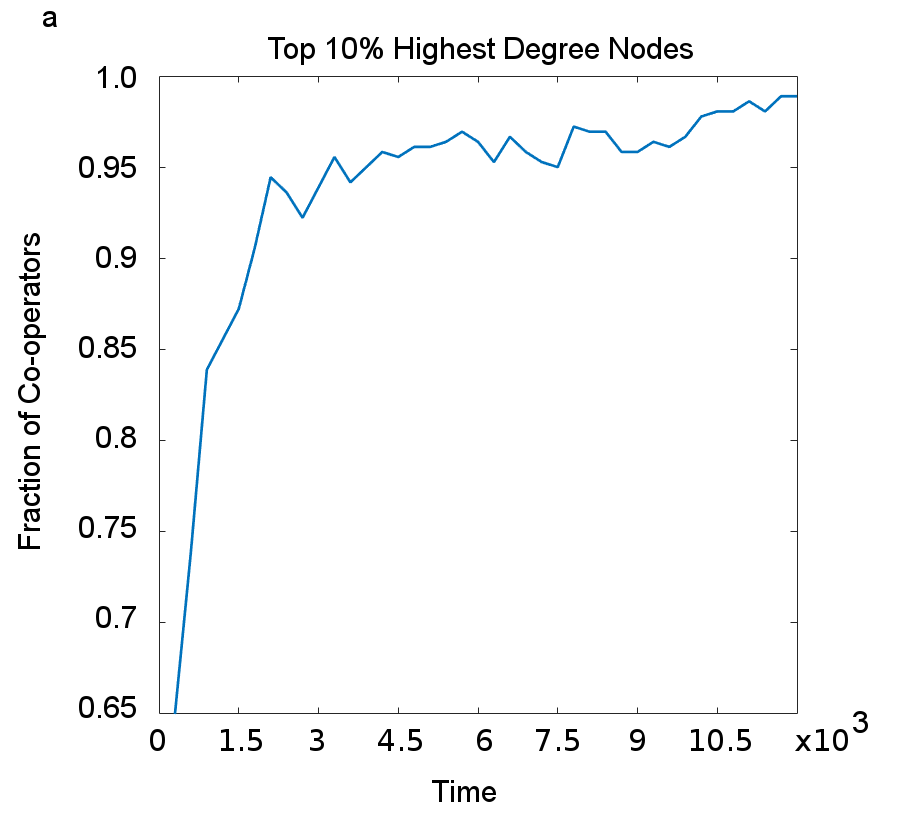}
\includegraphics[scale=0.45]{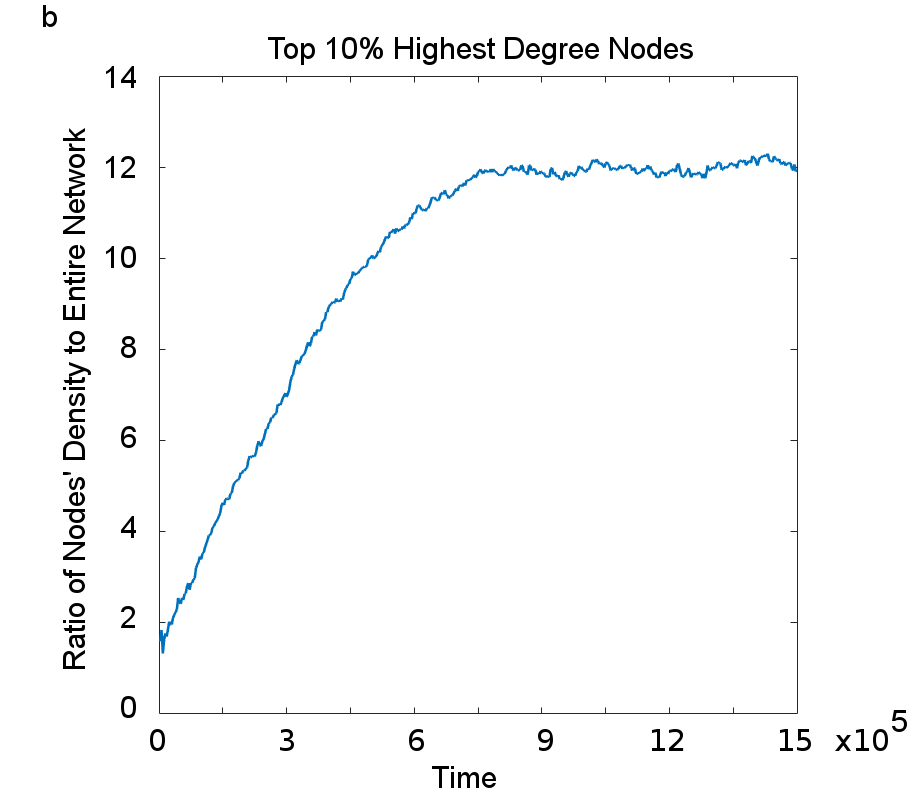}
\end{center}
\caption{}
\end{figure}

\begin{figure}[!ht]
\begin{center}
\includegraphics[scale=0.6]{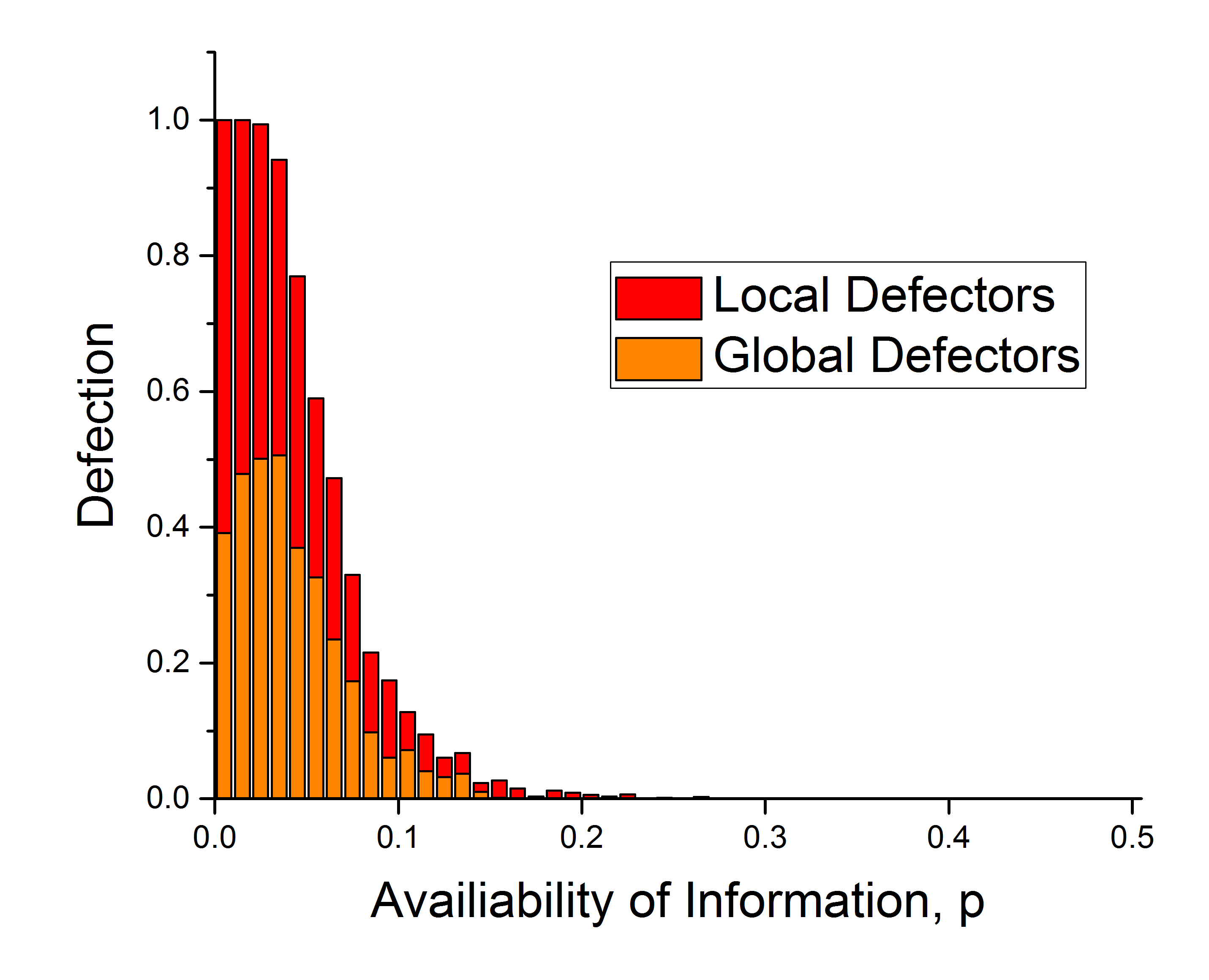}
\end{center}
\caption{}
\end{figure}

\begin{figure}[!ht]
\begin{center}
\includegraphics[scale=1]{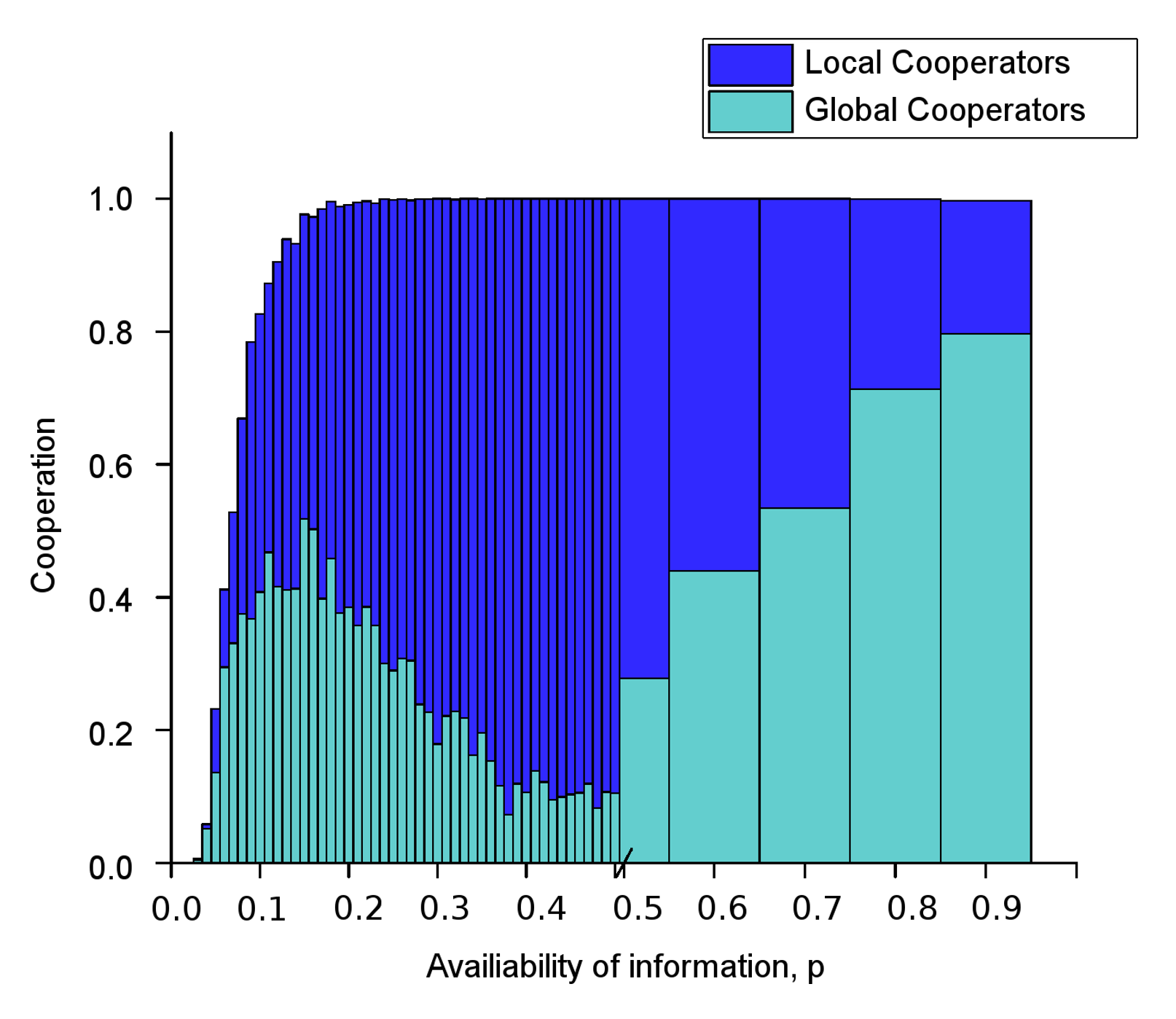}
\end{center}
\caption{}
\end{figure}

\begin{figure}[!ht]
\begin{center}
\includegraphics[scale=0.45]{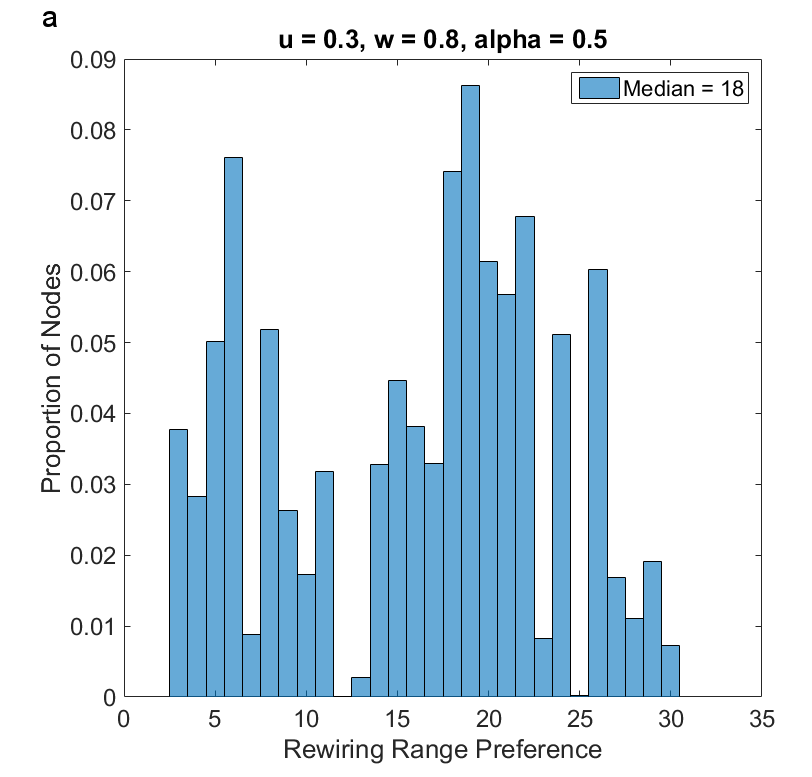}
\includegraphics[scale=0.45]{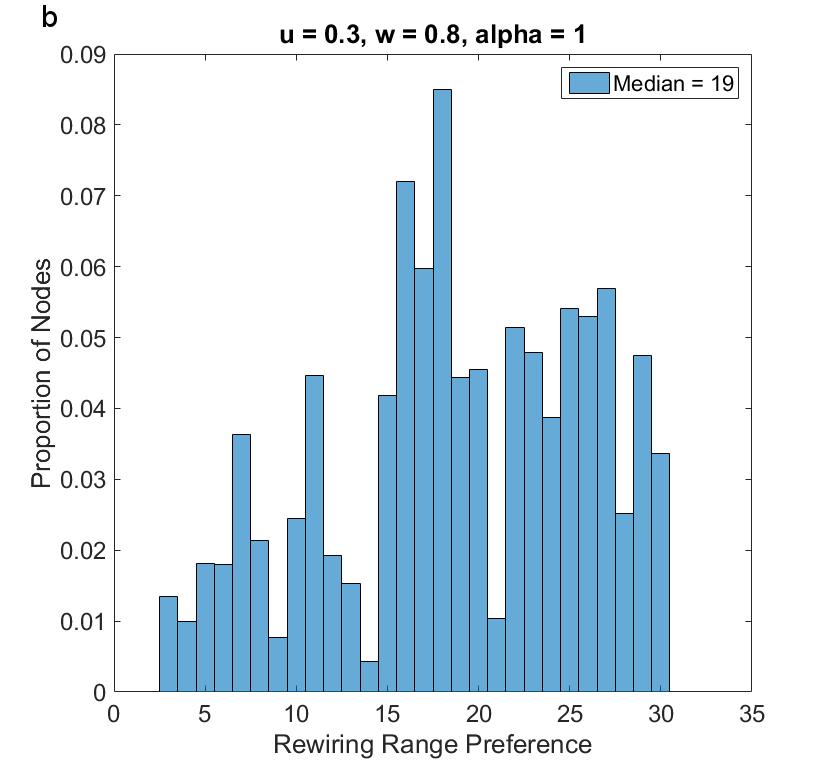}
\end{center}
\caption{}
\end{figure}

\begin{figure}[!ht]
\begin{center}
\includegraphics[scale=0.5]{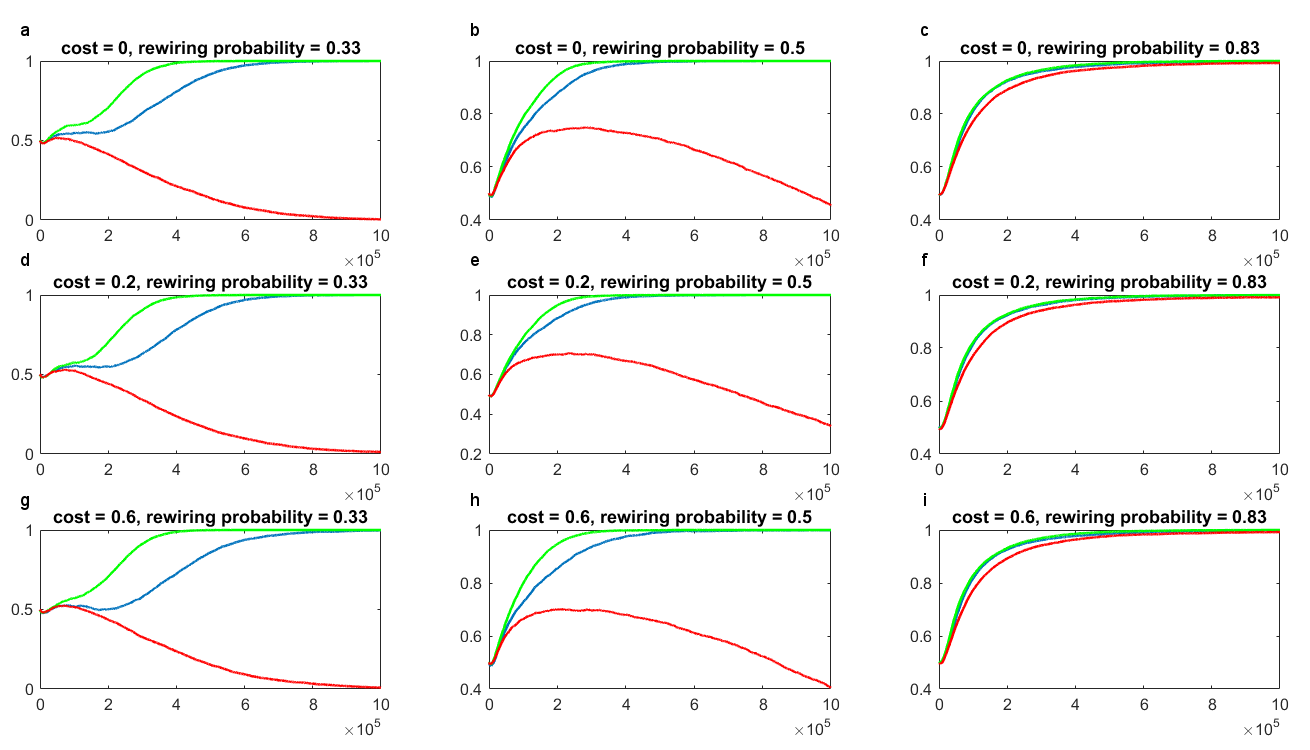}
\end{center}
\caption{}
\end{figure}

\begin{figure}[!ht]
\begin{center}
\includegraphics[scale=0.5]{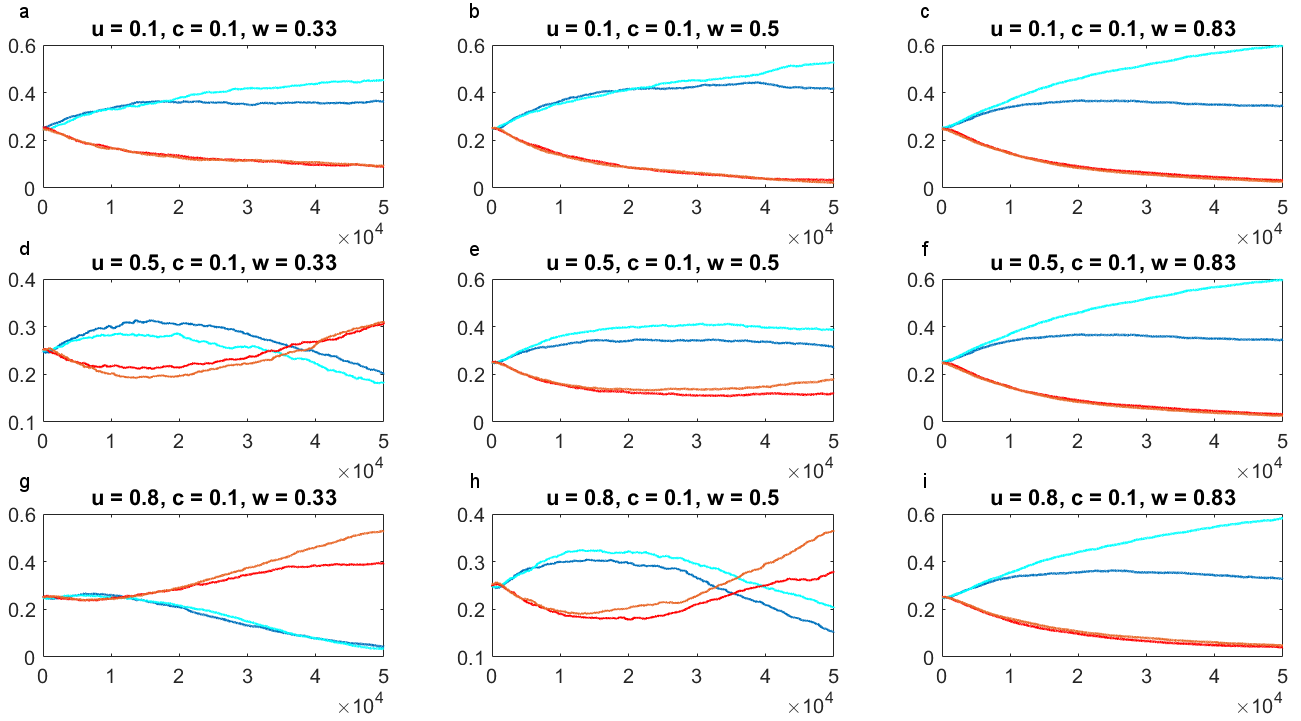}
\end{center}
\caption{}
\end{figure}

\begin{figure}[!ht]
\begin{center}
\includegraphics[scale=0.5]{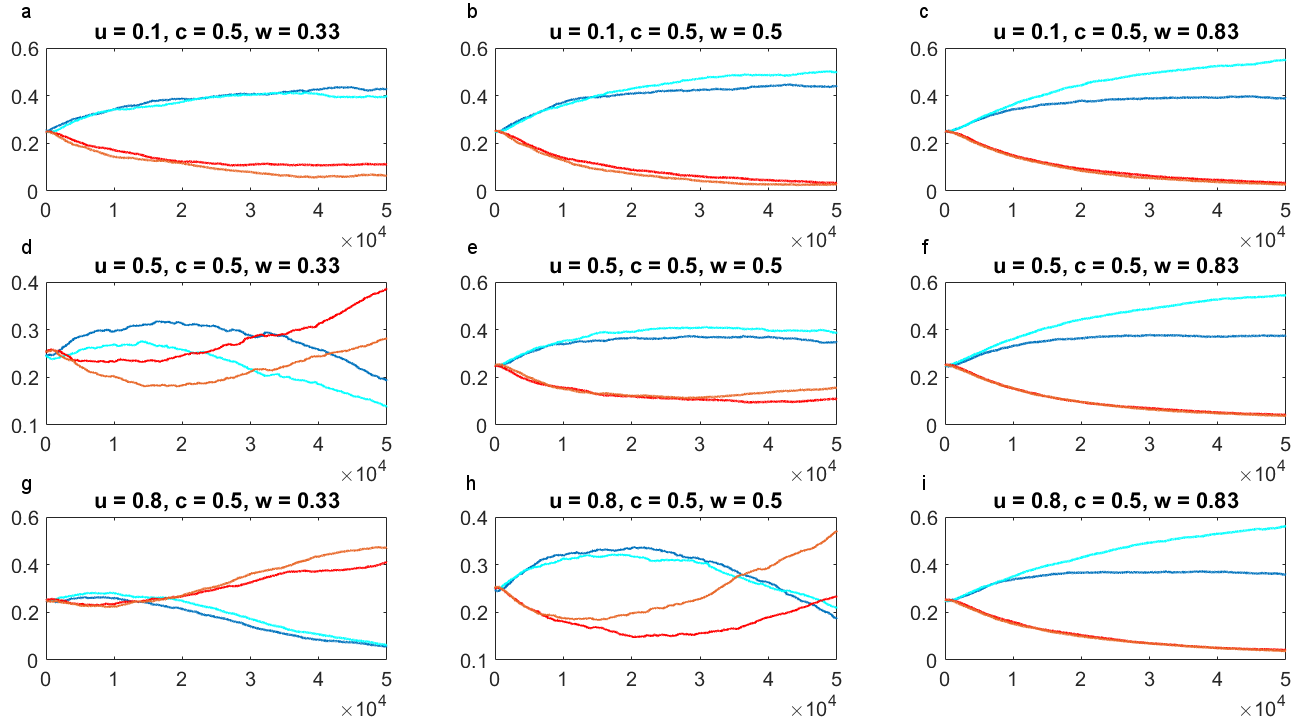}
\end{center}
\caption{}
\end{figure}

\begin{figure}[!ht]
\begin{center}
\includegraphics[scale=0.4]{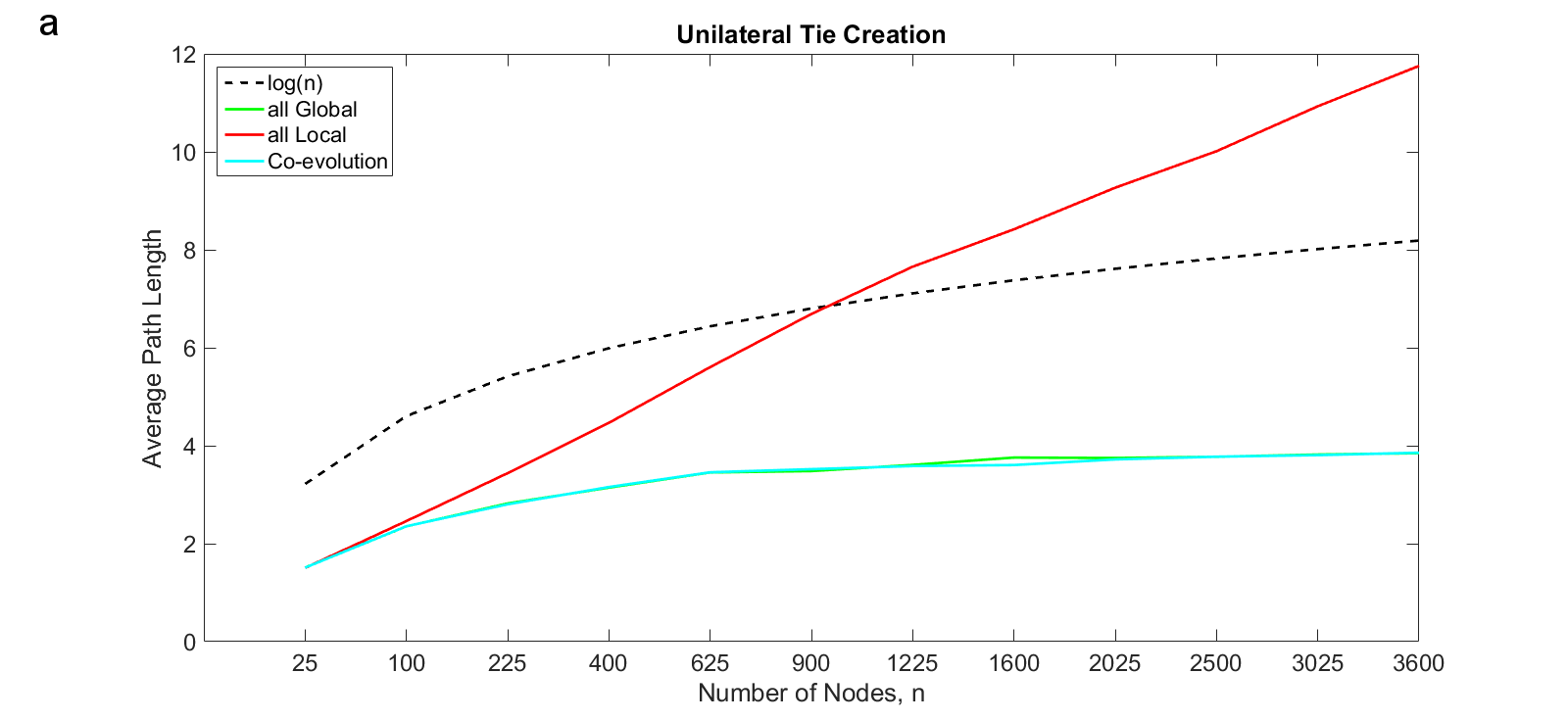}
\includegraphics[scale=0.4]{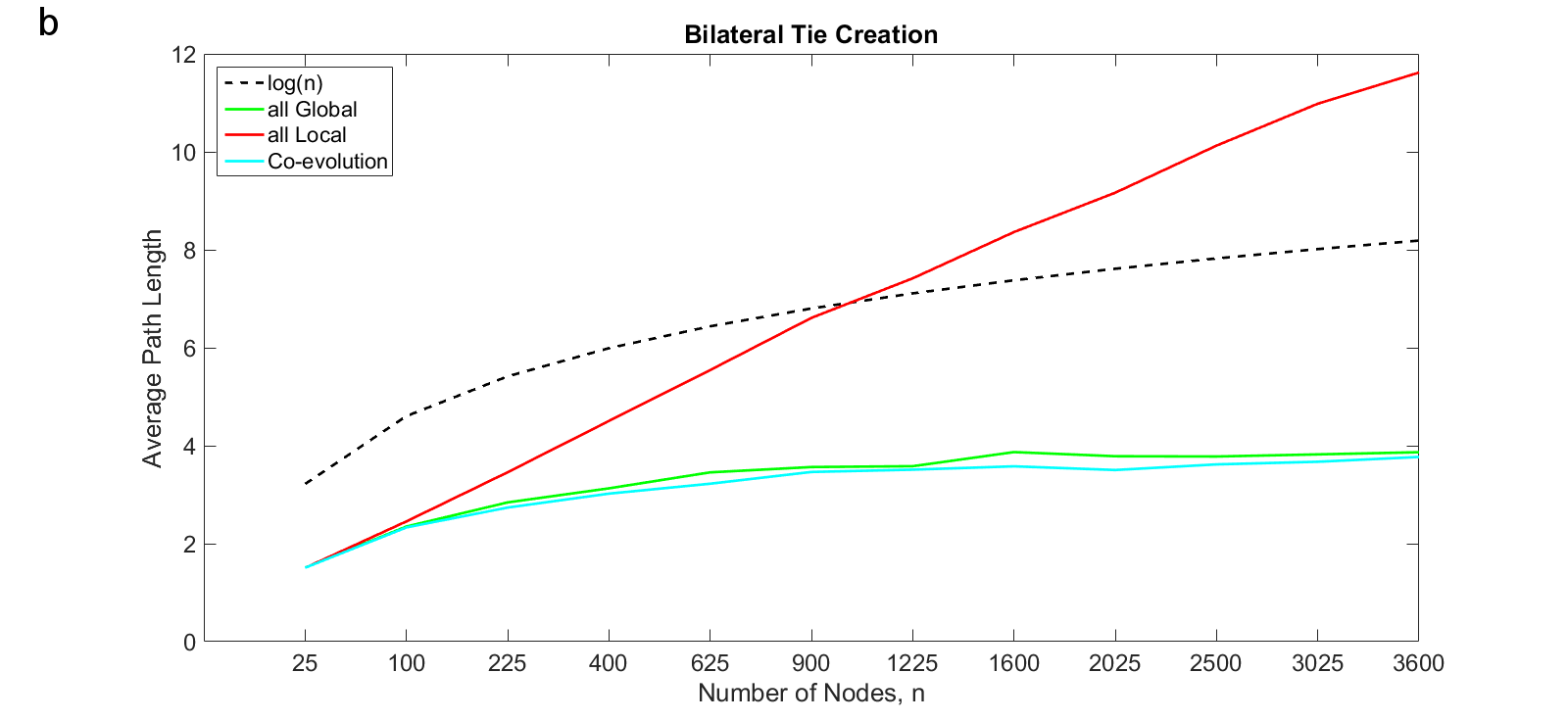}
\end{center}
\caption{}
\end{figure}

\begin{figure}[!ht]
\begin{center}
\includegraphics[scale=0.65]{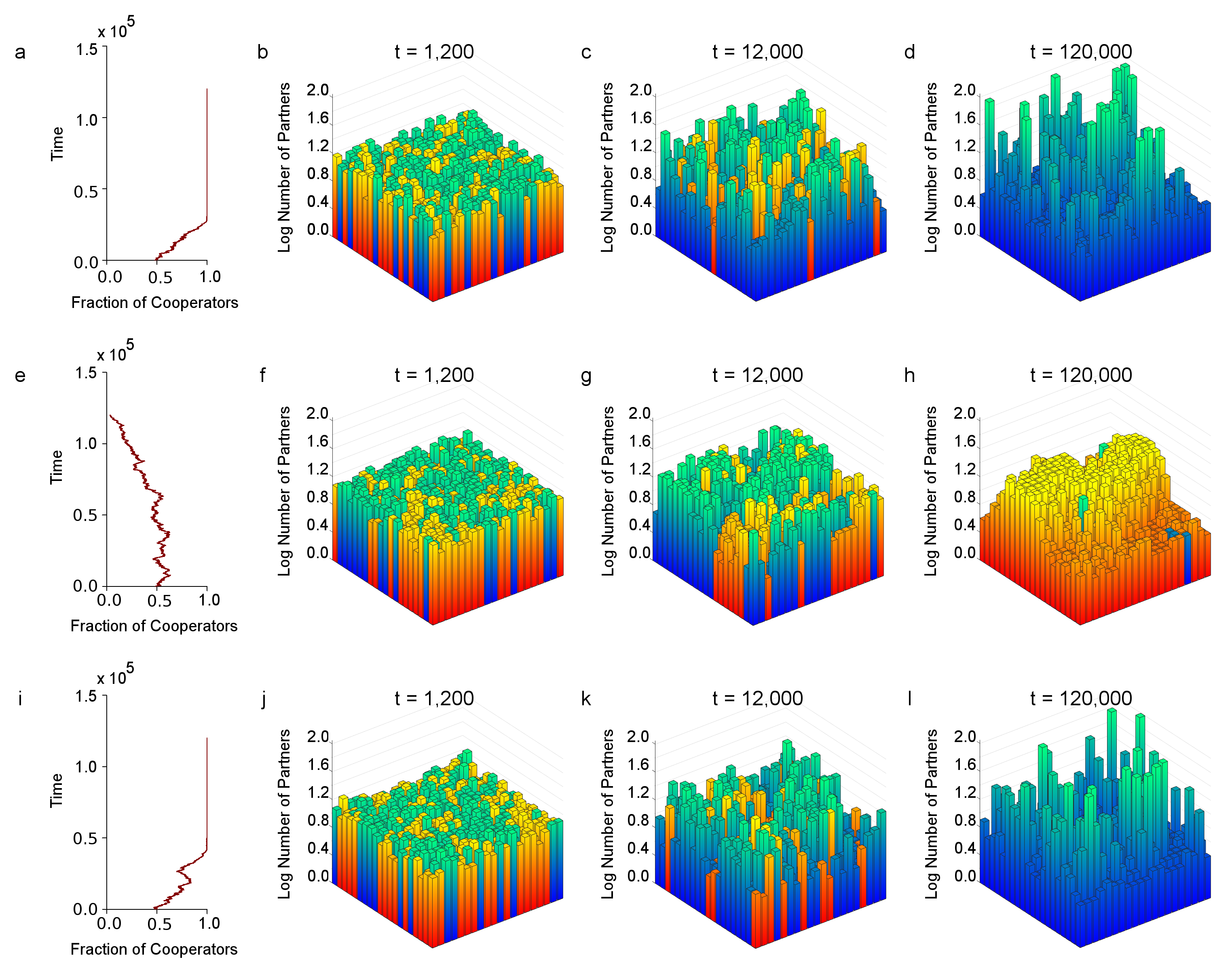}
\end{center}
\caption{}
\end{figure}

\begin{figure}[!ht]
\begin{center}
\includegraphics[scale=0.7]{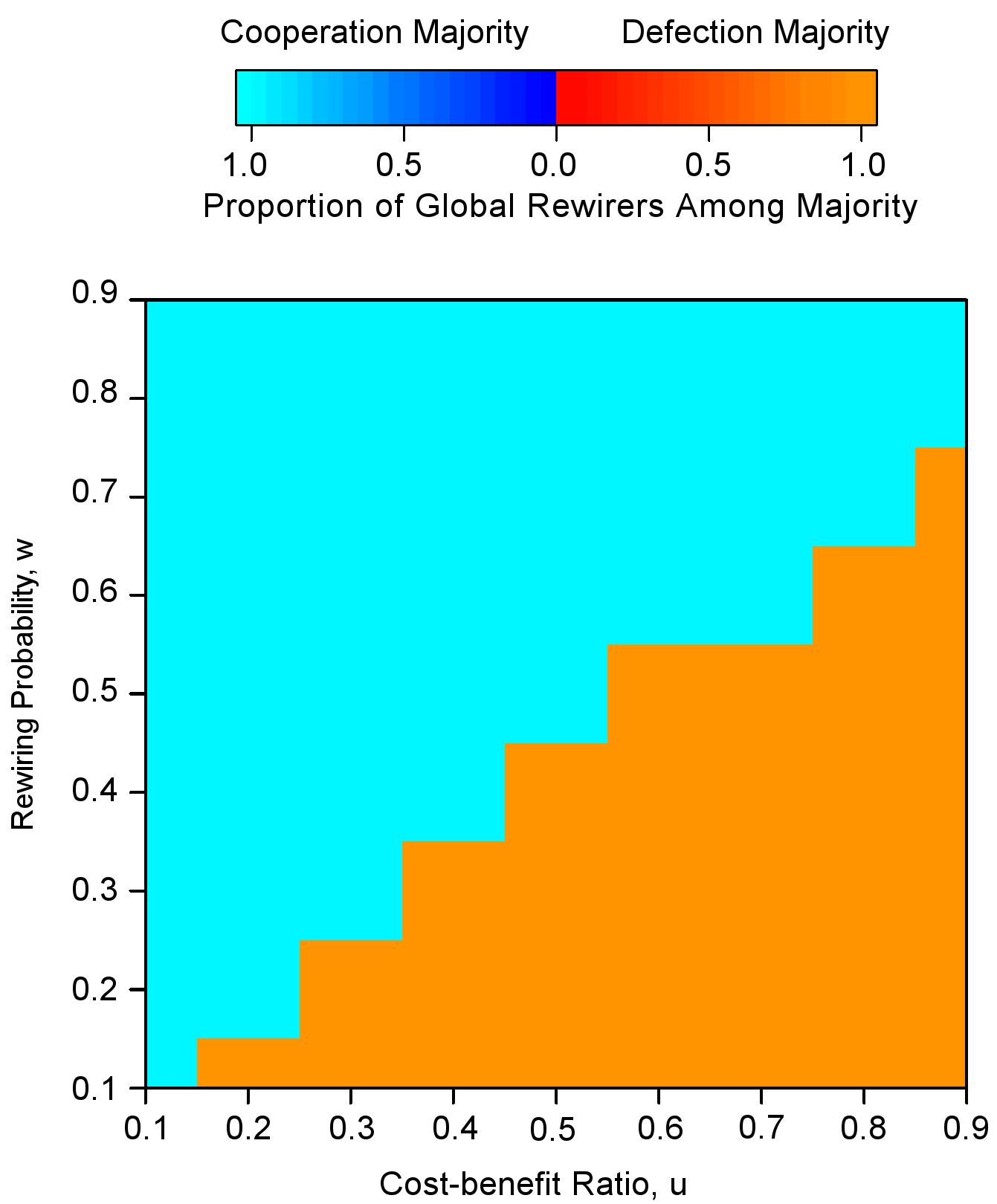}
\end{center}
\caption{}
\end{figure}

\begin{figure}[!ht]
\begin{center}
\includegraphics[scale=1]{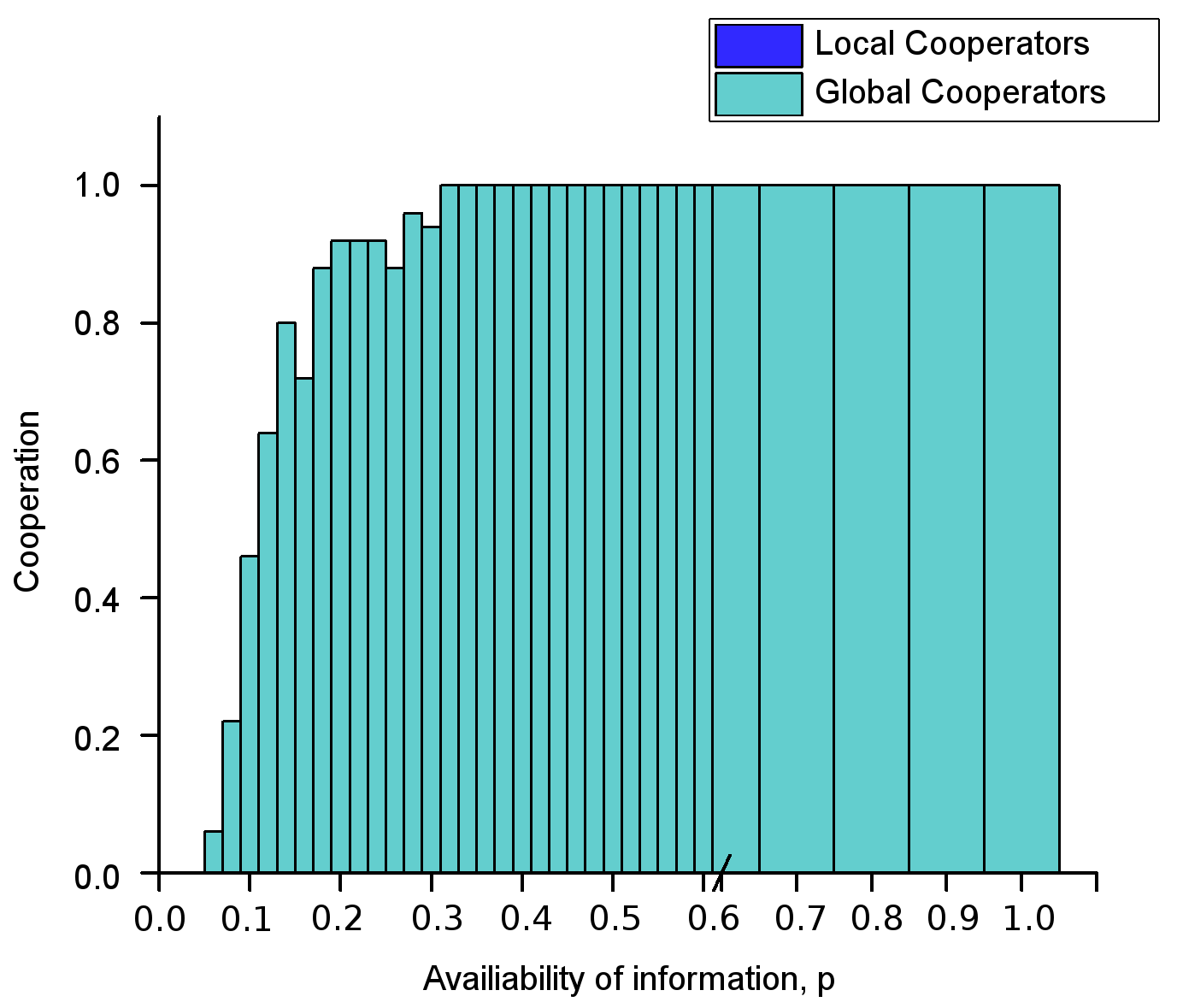}
\end{center}
\caption{}
\end{figure}

\end{document}